\documentclass[aps, prl, twocolumn, superscriptaddress]{revtex4-1}
\usepackage{amsmath}
\usepackage{amssymb}
\usepackage{graphicx}
\usepackage{color}

\newcommand{\ef}{\varepsilon_{\rm F}}
\newcommand{\kb}{k_{\rm B}}
\newcommand{\vg}{V_{\rm g}}
\newcommand{\vcn}{V_{\rm CN}}

\begin{document}

\title{Semiconducting-to-metallic photoconductivity crossover and temperature-dependent Drude weight in graphene}
\author{A. J. Frenzel}
\affiliation{Department of Physics, Massachusetts Institute of Technology, Cambridge, Massachusetts 02139, USA.}
\affiliation{Department of Physics, Harvard University, Cambridge, Massachusetts 02138, USA.}
\author{C. H. Lui}
\affiliation{Department of Physics, Massachusetts Institute of Technology, Cambridge, Massachusetts 02139, USA.}
\author{Y. C. Shin}
\affiliation{Research Laboratory for Electronics, Massachusetts Institute of Technology, Cambridge, Massachusetts 02139, USA.}
\author{J. Kong}
\affiliation{Research Laboratory for Electronics, Massachusetts Institute of Technology, Cambridge, Massachusetts 02139, USA.}
\author{N. Gedik}
\email{gedik@mit.edu}
\affiliation{Department of Physics, Massachusetts Institute of Technology, Cambridge, Massachusetts 02139, USA.}

\date{\today}

\begin{abstract}
We investigated the transient photoconductivity of graphene at various gate-tuned carrier densities by optical-pump terahertz-probe spectroscopy.  We demonstrated that graphene exhibits semiconducting positive photoconductivity near zero carrier density, which crosses over to metallic negative photoconductivity at high carrier density.  Our observations are accounted for by considering the interplay between photo-induced changes of both the Drude weight and the carrier scattering rate. Notably, we observed multiple sign changes in the temporal photoconductivity dynamics at low carrier density. This behavior reflects the non-monotonic temperature dependence of the Drude weight, a unique property of massless Dirac fermions.  
\end{abstract}

\maketitle

Charge carriers in graphene mimic two-dimensional massless Dirac fermions with linear energy dispersion, resulting in unique optical and electronic properties \cite{CastroNeto2009,*DasSarma2011}. They exhibit high mobility and strong interaction with electromagnetic radiation over a broad frequency range \cite{Mak2012}. Interband transitions in graphene give rise to pronounced optical absorption in the mid-infrared to visible spectral range, where the optical conductivity is close to a universal value $\sigma_0 = \pi e^2/2h$ (Ref. \onlinecite{Mak2008,*Li2008,*Wang2008}). Free-carrier intraband transitions, on the other hand, give rise to low-frequency absorption, which varies significantly with charge density and can cause strong extinction of light in the high density limit \cite{Horng2011,*Ren2012}.  In addition to this density dependence, the massless Dirac particles in graphene are predicted to exhibit a distinctive non-monotonic temperature dependence of the intraband absorption strength, or Drude weight, due to their linear dispersion \cite{Muller2009,Gusynin2009}. This behavior contrasts with the temperature-independent Drude weight expected in conventional systems of massive particles with parabolic dispersion \cite{Gusynin2007}.  While the unique behavior of the Drude weight in graphene has been considered theoretically, a clear experimental signature is still lacking. 

The intrinsic properties of Drude absorption in graphene can be revealed by studying its dynamical response to photoexcitation.  In particular, optical-pump terahertz-probe spectroscopy provides access to a wide transient temperature range via pulsed optical excitation, and allows measurement of the dynamical Drude conductivity by a time-delayed terahertz (THz) probe pulse \cite{Ulbricht2011}. This technique has been applied to study transient photoconductivity in graphene, but conflicting results have been reported \cite{Kampfrath2005,Jnawali2013,Frenzel2013,Docherty2012,Tielrooij2013,Choi2009,*Strait2011,*Winnerl2011,*Kim2013b}. Positive photoconductivity was observed in epitaxial graphene on SiC (Ref. \onlinecite{Choi2009}), while negative photoconductivity was seen in graphene grown by chemical vapor deposition (CVD; Refs. \onlinecite{Jnawali2013,Frenzel2013,Docherty2012,Tielrooij2013}). The opposite behavior in these samples has been argued to arise from their different charge densities. Gate tunable carrier density in a single sample promises to resolve these conflicts and allow observation of unique Dirac fermion behavior. 
  
In this Letter, we present an investigation of the Drude absorption dynamics in graphene over a wide range of carrier density and temperature. Using optical-pump THz-probe spectroscopy, we drove the carriers to high transient temperature and probed the Drude absorption of the hot carriers as they relaxed to equilibrium. By adjusting the pump-probe delay, we were able to observe the change of Drude absorption over a broad range of transient temperature. At low charge density, we observed complicated temporal dynamics of the THz Drude response, where the sign of photoconductivity changed multiple times as the carrier temperature increased and decreased.  This is a signature of the non-monotonic temperature dependence of the Drude weight in graphene, an intrinsic characteristic of massless Dirac fermions.  Additionally, we observed the carrier-density dependence of the dynamical Drude response through electrostatic gating of our samples. Near the charge neutrality point, our samples exhibited positive (semiconductor) photoconductivity, due to thermally excited electron-hole pairs after photoexcitation.  At high charge density, however, graphene exhibited negative (metallic) photoconductivity because of an optically-induced decrease of the Drude weight, coupled with increased carrier scattering rate. The observed density-dependent photoconductivity provides a unifying framework for understanding previously reported positive photoconductivity in (undoped) epitaxial graphene and negative photoconductivity in ($p$-doped) CVD graphene.  By using the Drude model with an estimated temporal evolution of the hot carrier temperature, we were able to reproduce all the main features of our observations. 

\begin{figure}[t]
   \includegraphics{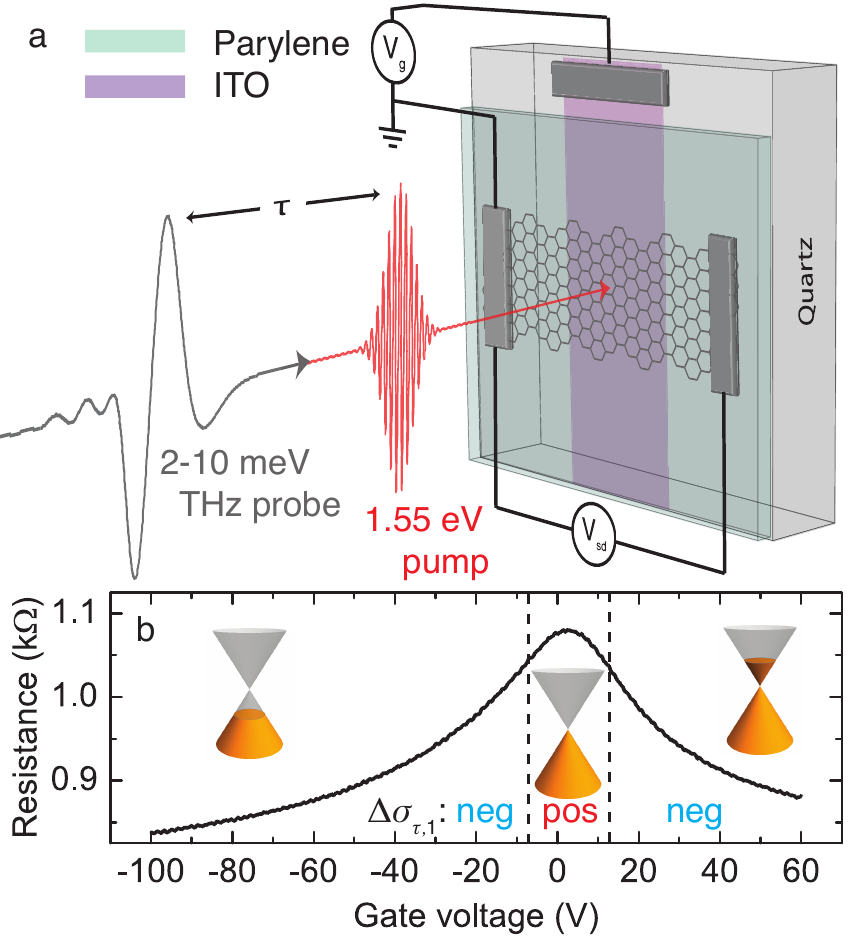}
   \caption{\label{schematic} (color online) (a) Schematic of transparent graphene device geometry and experimental method described in the text. (b) Two-terminal resistance of our device as a function of back gate voltage $\vg$. The charge neutrality point, corresponding to maximum resistance, is at $\vg = \vcn = 5$ V. Voltage ranges of positive and negative photoconductivity ($\Delta\sigma_{\tau,1}$) observed in our experiment are separated by dashed vertical lines.} 
\end{figure}

A key advance in our experiment is the fabrication of large-area gated graphene devices without a THz photoconductivity response from the substrate [Fig. \ref{schematic}(a)]. This is not possible for commonly used SiO$_2$/Si substrates, which produce large background signal in optical-pump THz-probe experiments \cite{Suzuki2011}. Here, we used $z$-cut crystalline quartz substrates and deposited 35-nm indium tin oxide (ITO) and 400-nm  parylene-C thin films as the back gate electrode and dielectric, respectively.  We experimentally confirmed that the back-gate structure had negligible pump-probe response (see Supplemental Material online, Ref. \onlinecite{SOM}). High-quality single-layer CVD graphene \cite{Li2009} sheets were transferred to our back-gate substrates.  Graphite-paint source and drain electrodes were attached to graphene with a separation of $\sim$5 mm. The devices exhibited excellent bipolar gating behavior with low unintentional doping [Fig. \ref{schematic}(b); gate voltage $\vg  = 5 $ V $\equiv \vcn $ at charge neutrality, corresponding to $n=3\times 10^{11}$ cm$^{-2}$, estimated from our device capacitance].

\begin{figure*}[h,t]
   \includegraphics{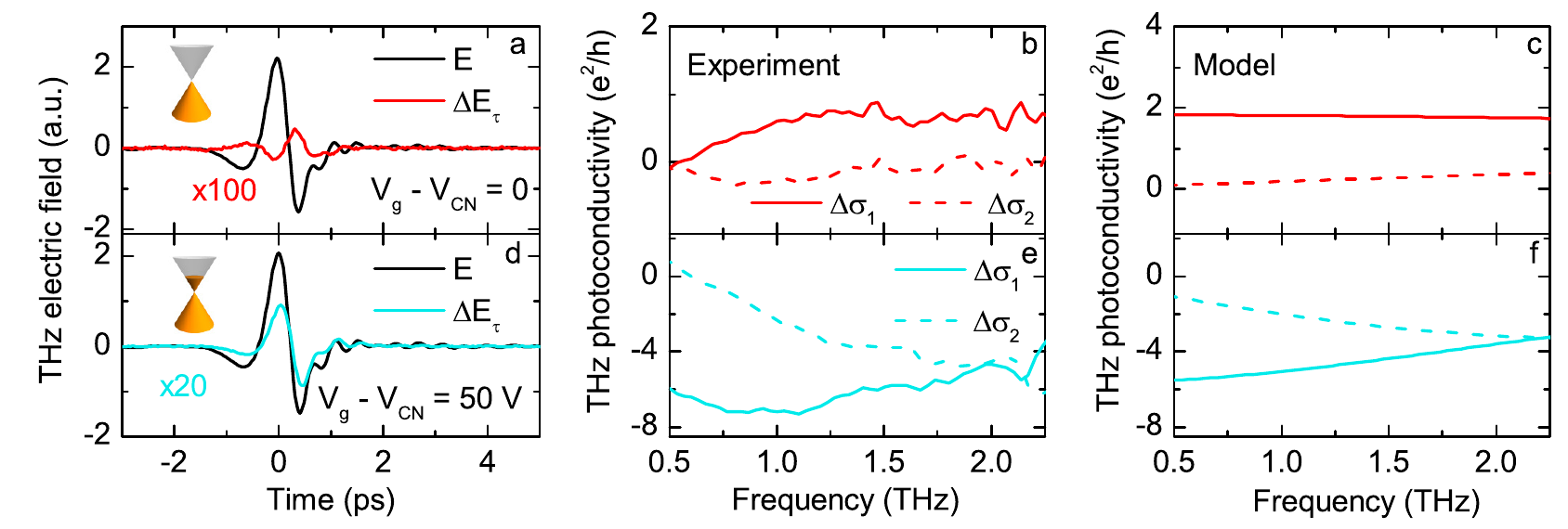}
   \centering
   \caption{\label{photoconductivity} (color online) (a) Measured THz electric field waveform transmitted through the sample in equilibrium (black line) and pump-induced change in transmitted THz electric field (red line) at $\tau = 1.5$ ps. Measurements were performed at room temperature in vacuum with the carrier density set to charge neutrality ($\vg = \vcn$) and incident pump fluence $\mathcal{F}=10$ $\mu$J/cm$^2$. (b) Real ($\Delta\sigma_1$, solid line) and imaginary ($\Delta\sigma_2$, dashed line) parts of the transient THz photoconductivity extracted from the data in (a). (c) Theoretical simulations of the photoconductivity spectra under the same conditions as (a-b) using the Drude model described in the text. (d-f) Experimental data and simulations as in (a-c), but at gate voltage +50 V from the charge neutrality point (electron density $n \approx 3 \times 10^{12}$ cm$^{-2}$).}
\end{figure*}

The graphene devices, investigated at room temperature in high vacuum ($P<10^{-5}$ Torr), were optically excited with 100 fs laser pulses at 1.55 eV photon energy generated using a 5 kHz amplified Ti:sapphire laser system. The transient photoconductivity was probed by measuring the complex transmission coefficient of picosecond THz pulses (photon energy 2-10 meV) with controllable time delay $\tau$ [Fig. \ref{schematic}(a)].  To reduce experimental errors associated with laser drift, we simultaneously measured the electric field waveform $E(t)$ of the THz pulse transmitted through the sample in equilibrium and the optical pump-induced change of the transmitted field $\Delta E_\tau (t)$ via electro-optic sampling \cite{Werley2011,SOM}.  In our experiments, the ratio $-\Delta E_\tau/E_0$ (referred to as ``differential field") approximately represents the photoconductivity $\Delta\sigma_{\tau,1}$ (Refs. \onlinecite{Kampfrath2005,Jnawali2013,Frenzel2013,SOM}).

Pump-probe measurements with incident pump fluence $\mathcal{F}=10 \, \mu$J/cm$^2$ and pump-probe delay $\tau = 1.5 $ ps reveal that the sign of the photoconductivity can be tuned from positive at charge neutrality to negative at a moderate carrier density [Fig. \ref{photoconductivity}]. The measured $\Delta E_\tau (t)$ at charge neutrality ($\vg = \vcn$) in panel (a) is everywhere opposite in sign to $E(t)$, reflecting a photo-induced increase in absorption. The photoconductivity $\Delta\sigma_\tau(\omega) = \Delta \sigma_{\tau,1} + i\Delta \sigma_{\tau,2}$, calculated from the waveforms in (a) and taking the device geometry into account, shows a positive real part [Fig. \ref{photoconductivity}(b)]. In sharp contrast, $\Delta E_\tau (t)$ has the same profile and sign as $E(t)$ when $\vg = \vcn + 50$ V ($n\approx 3\times 10^{12}$ cm$^{-2}$), implying a photo-induced decrease in absorption [Fig. \ref{photoconductivity}(d)]. As expected, the absorptive part of the photoconductivity, $\Delta \sigma_{\tau,1}$, is negative in this case [Fig. \ref{photoconductivity}(e)].

\begin{figure}[tb]
   \includegraphics{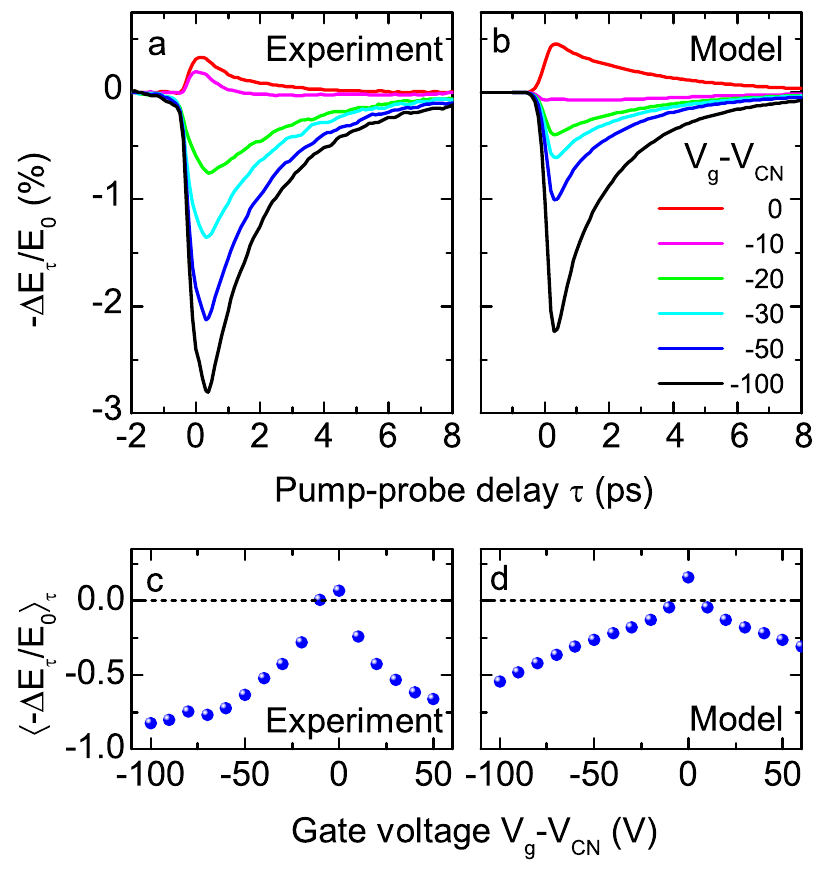}
   \caption{\label{dynamics} (color online) (a) Measured temporal evolution of the negative change in transmitted field (proportional to the differential conductivity), measured at the peak of the signal in Figs. \ref{photoconductivity}(a,d), at different gate voltages. Measurements were performed at room temperature in vacuum with incident fluence $\mathcal{F}=3$ $\mu$J/cm$^2$. (b) Theoretical simulation of the THz dynamics in (a), calculated using the model described in the text. (c) Mean of $-\Delta E(t=0)_\tau/E_0(t=0)$ from $\tau =$ -1 ps to 8 ps, as a function of gate voltage.  (d) Simulation of the data in (c).}
\end{figure}

To further investigate the mechanism driving the observed crossover, we measured the temporal dynamics of $\Delta \sigma_{\tau,1}$ at various carrier densities. Fig. \ref{dynamics}(a) displays the ratio $-\Delta E_\tau (t)/E(t)$ at fixed $t=0$ [Figs. \ref{photoconductivity}(a,d)] for gate voltages between -100 V and +50 V from $\vcn$. The relaxation time of the signal was approximately 2 ps, with no systematic dependence on carrier density. To more directly show the carrier density dependence, we plot the average value of the differential field, $\left<-\Delta E_\tau/E_0\right>_\tau$, as a function of gate voltage in Fig. \ref{dynamics}(c). These data show that the overall photoconductivity signal quickly changes from positive at charge neutrality to negative at moderate charge density. 

The opposite sign of $\Delta\sigma_{\tau,1}$ observed in graphene at the limits of charge neutrality and high charge density can be qualitatively understood by considering the interplay between photo-induced changes of carrier population and scattering rate (The more complicated behavior at low charge density is addressed in detail below).  Photoexcited carriers in graphene are known to thermalize rapidly within a few 10s of fs \cite{Lui2010, *Breusing2011, *Brida2013, *Johannsen2013, *Gierz2013}. Therefore, due to the 100 fs temporal resolution in our experiment, the charge carriers can be well described by a thermal distribution at temperature $T_{\rm e}$ for all pump-probe delay times $\tau$.  For graphene near the charge neutrality point, an increase of carrier temperature promotes the free-carrier population and thus enhances absorption. Such behavior mimics that observed in epitaxial graphene \cite{Choi2009,Strait2011} and other semiconductors \cite{Ulbricht2011}, where the optically generated electron-hole pairs increase the infrared absorption.  For graphene with high carrier density, laser-induced carrier heating only modifies the carrier distribution near the Fermi level, without changing the total carrier density.  The carrier scattering rate, however, increases due to an enlarged phase space for scattering and the presence of hot optical phonons after optical excitation.  This causes a reduction of free-carrier absorption, a behavior analogous to that in metals and observed in $p$-doped CVD graphene \cite{Tielrooij2013,Frenzel2013,Docherty2012,Jnawali2013}. 

\begin{figure}[tb]
   \includegraphics{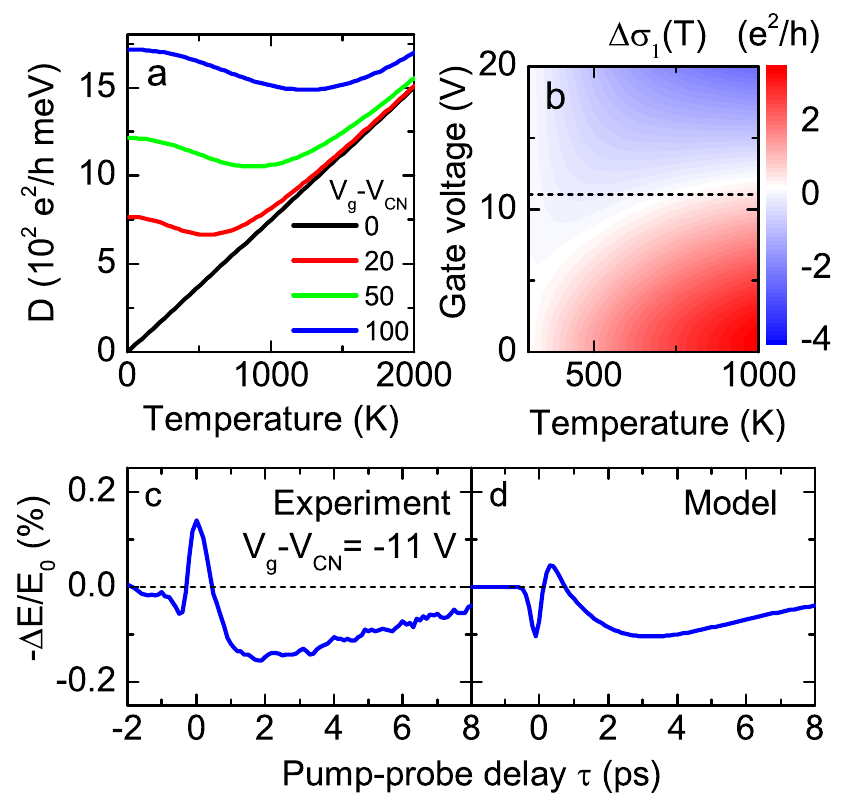}
   \caption{\label{simulation} (color online). (a) Temperature-dependent Drude weight [Eq. \eqref{drudeweight}] at different gate voltages. (b) Calculated change in conductivity $\Delta\sigma_1(T)$ at $\omega/2\pi=1$ THz, for different carrier densities and temperatures, relative to its value at $T = 300$ K. Temperature dependence of both Drude weight and scattering rate were taken into account. (c) Measured temporal dynamics of $-\Delta E_\tau/E_0$ at gate voltage $\vg = \vcn-11$ V [dashed line in (b)]. Incident fluence $\mathcal{F} = 10$ $\mu$J/cm$^2$. (d) Simulation of the temporal dynamics shown in (c), obtained using the model described in the text.}
\end{figure}

For a more thorough and quantitative analysis at all charge densities, we consider the Drude model for free carrier conductivity in graphene \cite{Gusynin2009,Horng2011,Strait2011,Frenzel2013,Jnawali2013},
\begin{equation}
\label{sigma}
\sigma(\omega) = \frac{D}{\pi(\Gamma-i\omega)}.
\end{equation}
Here, $\Gamma$ is the transport scattering rate and $D$ is the Drude weight, which quantifies the oscillator strength of free-carrier absorption. In a metal or semiconductor with parabolic dispersion, $D=\pi n e^2 /m$, independent of temperature \cite{SOM}. In graphene and other 2D systems with linear dispersion, however, $D$ exhibits a distinctive carrier temperature dependence \cite{Gusynin2009,Muller2009,Wagner2014,SOM}:
\begin{equation}
\label{drudeweight}
D(T_{\rm e})=\frac{2e^2}{\hbar^2}\kb T_{\rm e}\ln\left[2\cosh\left(\frac{\mu(T_{\rm e})}{2\kb T_{\rm e}}\right)\right].
\end{equation}
Fig. \ref{simulation}(a) shows that $D(T_{\rm e})$ increases linearly with temperature when $\kb T_{\rm e}\gg \ef $, and approaches $(e^2/\hbar^2)\mu\propto\sqrt{|n|}$ for electronic temperatures $\kb T_{\rm e}\ll \ef $. 

While $D$ only depends on the electron temperature $T_{\rm e}$, the transport scattering rate $\Gamma$ depends on $\ef $, $T_{\rm e}$, and the phonon temperature $T_{\rm ph}$ differently for different scattering mechanisms \cite{Hwang2008,*Fratini2008,*Hwang2009}. Optical excitation can therefore alter the THz conductivity [Eq. \eqref{sigma}] through the dependence of $D$ and $\Gamma$ on $T_{\rm e}$ and $T_{\rm ph}$. In our sample, we expect that charged impurities and hot optical phonons dominate the scattering rate \cite{Perebeinos2010,Hwang2008,Malard2013,SOM}. We calculated the temperature- and density-dependent change in conductivity relevant to our experiment, $\Delta\sigma_1(T_{\rm e}) = \sigma_1(T_{\rm e}) - \sigma_1(300$ K), at representative frequency $\omega/2\pi = 1$ THz, as shown in Fig. \ref{simulation}(b). For this calculation, we assumed that $T_{\rm ph} = T_{\rm e}$ (Refs. \onlinecite{Kampfrath2005,Lui2010}). The color plot shows that $\Delta\sigma_1(T_{\rm e})$ is positive (red area) near charge neutrality ($\vg < 5$ V), but becomes negative (blue area) at high carrier density ($\vg > 15$ V), as anticipated from the qualitative discussion above. 

We can convert the calculated $\Delta\sigma_1(T_{\rm e})$ to photoconductivity $\Delta\sigma_{\tau,1}$ by considering the temperature dynamics in graphene following photoexcitation. We estimated the transient temperature by comparison with previous publications \cite{Lui2010,Malard2013,Graham2013} and simulated the temporal photoconductivity dynamics. The time-dependent temperature was assumed to consist of two exponential decays with time constants $\tau_1 = 0.3$ ps and $\tau_2 = 3.1$ ps and a 200 fs rise time \cite{Malard2013,Graham2013}. The maximum estimated temperature was $\sim$500 K for 3 $\mu$J/cm$^2$ incident fluence and $\sim$800 K for 10 $\mu$J/cm$^2$ (Ref. \onlinecite{SOM}). Our simulated photoconductivity spectra, shown in Figs. 2(c,f), capture the main features of the experimental data [Figs. 2(a,d)]. Based on these spectra, we also calculated the expected $-\Delta E_\tau/E_0$, shown in Figs. 3(b,d). The good agreement between experiment and simulation demonstrates that our model correctly captures the relevant physics of transient photoconductivity in graphene.

We gain further insight into the transient photoconductivity dynamics in graphene by examining the transition regime between positive and negative photoconductivity [Fig. \ref{simulation}(c)]. The complicated temporal dynamics, characterized by multiple sign changes on a picosecond timescale, are due to the non-monotonic temperature dependence of the Drude weight. In graphene with finite carrier density, the Drude weight [Fig. \ref{simulation}(a)] initially drops as temperature increases, reaches a minimum at $\kb T_{\rm e} \approx 0.4 \,\ef$, then increases linearly at higher temperature \cite{Muller2009,SOM}. Correspondingly, $\Delta\sigma_1(T_{\rm e})$ at intermediate carrier density is initially negative, then becomes positive as temperature increases [dashed line in Fig. \ref{simulation}(b)]. In our pump-probe experiment, the optical pump pulse initially heats the carriers to high temperature, resulting in a sign change of $\Delta\sigma_{\tau,1}$ from negative to positive. As the carriers then cool, $\Delta\sigma_{\tau,1}$ changes sign to negative again before returning to zero at equilibrium. Our observed temporal photoconductivity dynamics [Fig. \ref{simulation}(c)], which follow this sequence of sign changes, explicitly reveal the non-monotonic temperature-dependence of the Drude weight in graphene. Using the same parameters as before, our model reproduces this behavior [Fig. \ref{simulation}(d)]. We note that these temporal dynamics do not appear in our simulation if we neglect the non-monotonic temperature dependence of $D(T_{\rm e})$ (Ref. \onlinecite{SOM}).

The unusual non-monotonic temperature dependence of the Drude weight can be understood by considering  the conservation of spectral weight of optical transitions \cite{Kuzmenko2008,Gusynin2009,Horng2011}. The optical absorption spectrum in graphene consists of two contributions: high-energy interband absorption and low-energy intraband absorption. Interband absorption in graphene with finite charge density shows an onset at photon energy $\hbar\omega = 2|\mu|$ due to Pauli blocking \cite{Mak2008,Gusynin2009,Horng2011}. When carriers are heated to moderate temperatures $\kb T_{\rm e} \ll \ef$, $\mu(T_{\rm e})$ decreases due to particle conservation \cite{Ashcroft1976}. This decreases the interband absorption threshold, increasing the interband spectral weight. To conserve total spectral weight, the intraband absorption must therefore decrease. When carrier temperatures become comparable to $\ef$, however, interband transitions are Pauli blocked by thermally excited carriers, reducing the spectral weight. This increases the intraband spectral weight, as has been observed in graphite \cite{Kuzmenko2008}. We emphasize that this behavior is due to the linear dispersion of charge carriers in graphene, and is absent in materials with parabolic dispersion.

In conclusion, we have studied the temperature- and density-dependent Drude conductivity in graphene through its dynamical response to pulsed photoexcitation.  We demonstrated that the transient photoconductivity behavior of graphene can be tuned continuously from semiconducting to metallic by varying the Fermi level from the charge neutrality point to either the electron or hole side.  This resolves the controversy between previous experiments which observed positive photoconductivity in epitaxial graphene and negative photoconductivity in CVD graphene.  By detailed simulation based on the Drude model, we found that photo-induced changes of both Drude weight and carrier scattering rate play important roles in the THz photoconductivity dynamics.  In particular, complicated temporal evolution of the photoconductivity at low charge density serves as a direct signature of the non-monotonic temperature dependence of the Drude weight in graphene, a unique property of systems with linear energy dispersion.

Note: upon completion of this work, we became aware of similar work by another group \cite{Shi2014}.

\begin{acknowledgments}
The authors acknowledge helpful discussions with J.C.W. Song and O. Khatib, and thank V. Fatemi, J.D. Sanchez-Yamagishi, and M.A. Smith for assistance with device fabrication. This work was supported by Department of Energy Office of Basic Energy Sciences Grant No. DE-SC0006423 (sample fabrication, experimental setup, and data acquisition) and STC Center for Integrated Quantum Materials, NSF grant DMR-1231319 (data analysis). A.J.F. acknowledges support from NSF GRFP. This work also made use of Harvard's Center for Nanoscale Systems (CNS), supported by the National Science Foundation under grant No. ECS-0335765, and the MIT Microsystems Technology Laboratory (MTL). 
\end{acknowledgments}

\end{document}